\documentclass[12pt]{article}
\usepackage{newtxtext,newtxmath}
\usepackage{upgreek}
\usepackage{graphicx}
\usepackage[letterpaper,margin=1in]{geometry}
\renewenvironment{abstract}
	{\quotation}
	{\endquotation}

\date{}

\makeatletter
\renewcommand{\fnum@figure}{\textbf{Figure \thefigure}}
\renewcommand{\fnum@table}{\textbf{Table \thetable}}
\makeatother

\usepackage{scicite}

\usepackage{url}

\newcommand\eaq {\text{e}^-_{\text{aq}} }

\newcommand{\uGy}{$\upmu$Gy}

\usepackage[version=4]{mhchem}

\def\scititle{
	Real-time tissue-equivalent measurement of individual clinical radiotherapy pulses
}
\title{\bfseries \boldmath \scititle}

\author{
	Fernanda~C.~Rodrigues-Machado$^{1\ast}$,
	Katherine~Szabo$^{1}$,\and
	Jingyi~Bian$^{2}$,
    Simon~Bernard$^{1}$,
    Tanner~Connell$^{2}$,\and
    Shirin~A.~Enger$^{2}$,
    Lilian~Childress$^{1}$,
    Jack~C.~Sankey$^{1\ast}$ \and
	\small$^{1}$Department of Physics, McGill University, Montréal \& H3A 2T8, Canada.\and
	\small$^{2}$Medical Physics Unit, McGill University, Montréal \& H4A 3J1, Canada.\and
	\small$^\ast$Corresponding authors. Emails: fernanda.rodriguesmachado\@mail.mcgill.ca, jack.sankey\@mcgill.ca
}

\begin{document} 

\maketitle

%%%%%%%%%%%%%%%% ABSTRACT

% SUBMISSION: Uncomment the boldface stuff
\begin{abstract} \bfseries \boldmath
We apply the precision tools of cavity-enhanced absorption sensing to clinical oncology, demonstrating a dosimeter paradigm in which a centimeter-scale volume of water serves as a tissue-equivalent sensing medium. Our proof-of-concept, all-optical scheme achieves real-time readout of clinical radiation pulses with a nominal single-pulse resolution of 90~\uGy. This demonstration paves the way toward fiber-integrated, micron-scale devices for \textit{in situ} universal absolute dosimetry during treatment.
\end{abstract}

%%%%%%%%%%%%%%%% INTRODUCTION

% The first paragraph of any Science paper does NOT have a heading
% Nor is it indented
\noindent
Accurate dosimetry is central to modern cancer radiotherapy, directly impacting tumor control, healthy tissue sparing, and ultimately the outcomes of $\sim$10~million patients each year ($\sim$50\% of all cancer patients worldwide~\cite{zhu2024}).
Despite major advances in beam delivery and treatment planning, \textit{in vivo}, real-time monitoring of the dose delivered locally during treatment has not become a standard part of clinical practice~\cite{mijnheer2013}.
Instead, clinical workflows rely on pre‑treatment beam characterization, calibrated external reference dosimeters, and strict patient positioning to maximize agreement between planned and delivered dose~\cite{podgorsak2016,chow2025}. This indirect or delayed approach precludes the detection and correction of dose fluctuations during treatment arising from beam instabilities, patient motion, or setup errors.

Existing dosimetry technologies also lack a compact, real‑time, absolute dosimeter whose active sensing medium is radiologically equivalent to human tissue. Most clinically deployed detectors use solid‑state or gas‑based materials whose response depends strongly on radiation type, energy, and geometry, necessitating correction factors that introduce systematic uncertainties~\cite{IAEA2024,gibbons2020}. In contrast, the absorbed dose to water represents a nearly ideal proxy for the relevant biological processes, and consequently serves as the primary (absolute) calibration reference in radiotherapy dosimetry; it is currently measured via careful calorimetry with high doses, and only at specialized (non-clinical) laboratories~\cite{podgorsak2016}. A high-speed detector sensitive to small doses delivered to water would therefore represent an ideal solution for both monitoring during clinical treatment and rapid local quality assurance~\cite{podgorsak2005}.

The need for such a detector is especially acute in the context of emerging treatment modalities. Ultrahigh dose rate (FLASH) radiotherapy, for example, has demonstrated remarkable selectivity in preclinical studies~\cite{favaudon2014}, yet its clinical translation is limited by the lack of reliable dosimeters~\cite{siddique2023,EURAMET2024}. Similar challenges arise in magnetic‑resonance‑guided radiotherapy~\cite{depooter2021,sarfehnia2025} and in particle‑based treatments~\cite{IAEA2024}, where conventional dosimeters may fail or require extensive corrections. A universal dosimeter operating consistently across all radiation types and dose rates would eliminate these barriers.

One promising proxy for absorbed dose to water is the hydrated electron~($\eaq$)~\cite{alizadeh2012,jay2025}, which exhibits broadband optical absorption in the visible spectrum~\cite{hart1962,torche2016}, and has a yield proportional to dose~\cite{spinks1990}. In principle, these properties enable all‑optical readout without the need for electrodes, charge collection, or material‑specific corrections. However, even with quantum‑limited laser sources and state‑of‑the‑art photodetectors, the absorption signal from clinical radiation pulses is so small that meter‑scale optical paths are required to achieve useful sensitivity. Large water tanks (or very high dose experiments) have suggested the promise of water as a universal sensing medium~\cite{fielden1968,megroureche2023,terfas2025,cao2026}, but their size and low sensitivity render them impractical for clinical deployment.

Recently, low‑loss dielectric mirror coatings have been shown to withstand extreme radiation exposure without degradation~\cite{rodrigues2024}, providing a pathway toward dramatically increased sensitivity and compactification. By enclosing the water volume within a high‑finesse Fabry--Pérot optical cavity, meters-long effective path lengths can be realized with devices that are many orders of magnitude smaller, in principle down to the micron scale at the tip of an optical fiber~\cite{hunger2010}. This would enable \textit{in situ} applications such as integration with existing intervention tools (the tip of a biopsy needle or catheter) or endoscopic probes during exposure.

Here we demonstrate a cavity‑enhanced, high‑speed, water‑based dosimeter capable of resolving individual pulses from a clinical linear accelerator with a sensitivity below 100~\uGy. This intermediate-scale prototype is 3~cm long, representing a 40-fold compactification. We characterize the device response to 6~MV, 10~MV, and 15~MV photon beams, directly observing the time-evolution of radiation-induced hydrated electron populations via their optical absorption, a real-time signal proportional to the absorbed dose to water. In addition to establishing the operating principles of this dosimetry paradigm, this device demonstrates the feasibility of absolute dosimetry for \textit{in vivo} readout and quality assurance in a clinical setting, and provides critical early guidance for further compactification to the sub-millimeter regime.

\section*{Single-Shot Readout}

\begin{figure} 
	\centering
	\includegraphics[width=.8\textwidth]{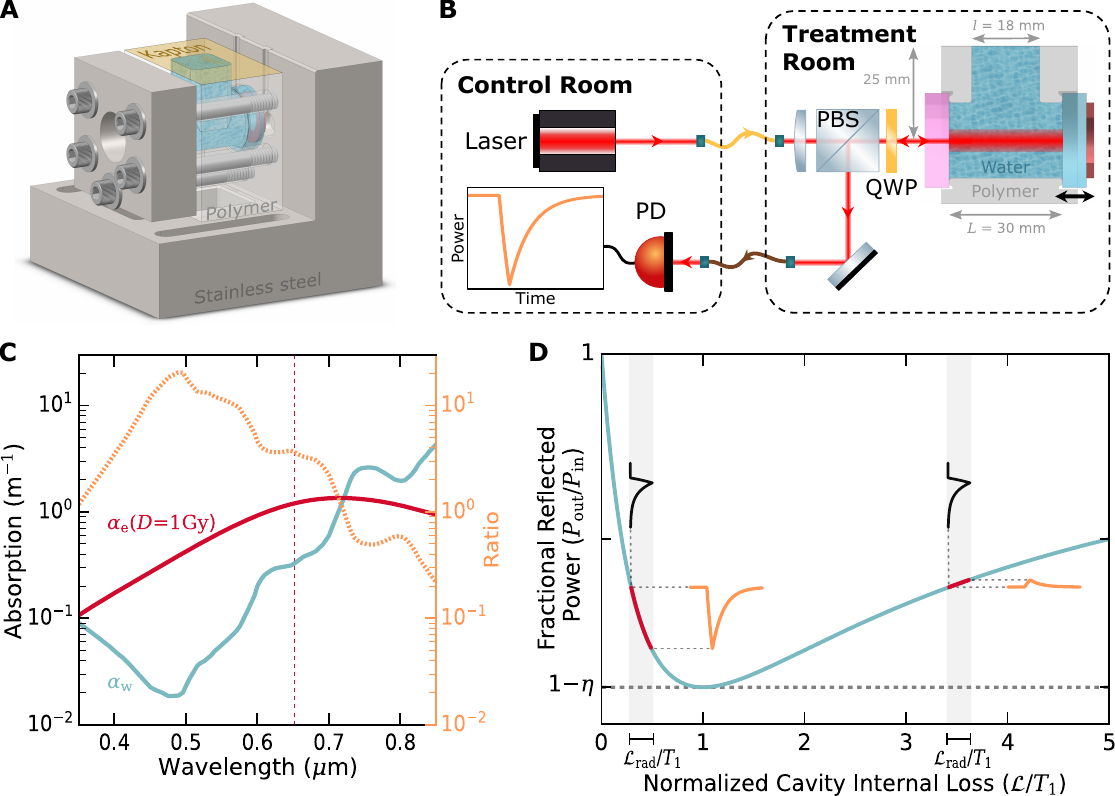} 
	\caption{\small
        \textbf{Optical setup and readout mechanism.}
		(\textbf{A})~Water-filled optical cavity, showing aqueous solution (blue) in a polymer basin (transparent), with cavity mirrors clamped (grey structure) at each end. Cavity length is adjusted by a piezoelectric ring transducer (red). Radiation from a clinical LINAC enters via a top aperture of $18\times20$~mm$^2$ that is sealed by two layers of 64-$\upmu$m-thick Kapton tape (yellow) to limit oxygen contamination.
        (\textbf{B})~Detection and readout. Laser light (652~nm) is coupled to a 20~m single-mode fiber (yellow) for delivery to the treatment room, where it is collimated and mode-matched to the TEM$_{00}$ cavity mode 25~mm below the water surface. A polarizing beam splitter (PBS) and quarter wave plate (QWP; 45$^\circ$~orientation) maximize the reflected light coupled into an multimode fiber (brown, 18~m) leading back to a photodetector (PD) in the control room. Water radiolysis generates light-absorbing hydrated electrons~($\eaq$), temporarily modifying the collected optical power, producing a $\sim$10-$\upmu$s transient (orange curve).
        (\textbf{C})~Wavelength-dependent optical absorption of hydrated electrons $\alpha_\text{e}$ (red~\cite{torche2016}; for a dose $D=1$~Gy) and background water absorption $\alpha_{\text{w}}$ (teal~\cite{segelstein1981}), with their ratio shown in orange. The dashed vertical line references our red (652~nm) laser. 
        (\textbf{D})~Dependence of fractional collected power ($P_\text{out}/P_\text{in}$) on cavity internal round-trip loss $\mathcal{L}$ normalized by input mirror transmission $T_1$ (teal curve). Hydrated electrons transiently increase $\mathcal{L}$ by $\mathcal{L}_\text{rad}$ (exaggerated black curves), producing a corresponding fluctuation (orange curves) according to the slope (red), changing sign depending on whether the cavity is overcoupled ($\mathcal{L}/T_1<1$, left) or undercoupled ($\mathcal{L}/T_1>1$, right). The mode-matching parameter $\eta$ sets the minimum value of $P_\text{out}/P_\text{in}$.
		}
	\label{fig:fig1}
\end{figure}

Figure~\ref{fig:fig1}A shows the prototype geometry, namely a 3-cm-long water-filled polymer basin with a horizontal cylindrical 2-cm-diameter through-hole (sealed by cavity mirrors) and an 18~mm~$\times$~20~mm vertical aperture through which radiation can pass. The cavity mirrors have nominal power reflectivity of 96\% (input) and 99.998\% (back) optimized for reflection-mode measurement. 
We fill the basin with an aqueous NaOH solution (pH~12.12) purged with argon gas to reduce dissolved oxygen and increase the observable $\eaq$ radiation yield~\cite{spinks1990,buxton1988,keene1962,butler1994}. The cavity's optical axis resides within the solution 25~mm below the surface, allowing production and build-up of secondary electrons prior to passage through the circulating light~\cite{podgorsak2016}. 

The cavity is driven by a 652~nm laser (Fig.~\ref{fig:fig1}B) that, along with all electronics, resides in the control room and is protected from stray radiation. A 20-m-long single-mode optical fiber carries input light to the treatment room, where the beam is collimated and coupled to the TEM$_{00}$ mode of the optical cavity. We utilize a polarizing beam splitter and quarter-wave plate to maximize collection of reflected power, which is coupled into a multimode fiber for delivery to a photodiode in the control room. 
The cavity resonance is locked to the laser using a Pound-Drever-Hall error signal~\cite{black2001} fed back to a ring piezoelectric transducer behind the back mirror to adjust cavity length.

The aqueous solution is exposed to $\sim$${4\ \upmu}$s irradiation pulses from a clinical linear accelerator (LINAC). Water radiolysis produces hydrated electrons ($\eaq$) through ionization of water molecules followed by rapid thermalization of the ejected electrons, resulting in a longer-lived population of $\eaq$ that can be optically excited by visible-wavelength photons~\cite{pizzochero2019}. The resulting $\eaq$ absorption (Fig.~\ref{fig:fig1}C, red curve~\cite{torche2016}) adds to the intrinsic absorption of water (teal curve~\cite{segelstein1981}), increasing the round-trip internal optical loss $\mathcal{L}$ of the cavity. 
While $\eaq$ absorption peaks around 715~nm, water absorption increases sharply in the near-IR region, so operating at shorter wavelengths increases the ratio between the two (dashed orange curve).
Due to the finite lifetime of the hydrated electron, incident radiation produces a transient in the optical absorption that qualitatively resembles the inset plot of Fig.~\ref{fig:fig1}B, with a characteristic build-up during irradiation and subsequent decay determined by the $\eaq$ lifetime ($\sim$10~$\upmu$s~\cite{megroureche2023}). 

Quantitatively, the fraction of incident laser power returning from a Fabry-P{\'e}rot cavity on resonance is~\cite{rodrigues2022}
    \begin{align}   \label{eq:R}
    R &\approx 
        \left(\frac{  \mathcal{L}_0 + \mathcal{L}_{\text{rad}}(t) -T_1 }
        { \mathcal{L}_0 + \mathcal{L}_{\text{rad}}(t) + T_1  }\right)^2 \ ,
    \end{align}
where $T_1$ is the (power) transmission coefficient of the input mirror, ${\mathcal{L}_0=T_2+2\delta+2\alpha_{\text{w}} L}$ is the nominal roundtrip ``internal'' optical losses of the cavity -- including (minuscule) transmission $T_2$ through the back mirror, mirror scattering and absorption $\delta$, and ``background'' absorption from the aqueous medium with coefficient $\alpha_\text{w}$ over the cavity length $L$ -- and $\mathcal{L}_{\text{rad}} = 2\alpha_\text{e}L$ is the \textit{extra} absorption due to radiation-induced hydrated electrons having absorption coefficient $\alpha_\text{e}(t)$ (time-dependence discussed below). The optical power $P_\text{out}$ reaching the photodetector is then given by 
    \begin{equation}\label{eq:Pout_and_eta}
    P_\text{out} \approx P_\text{in} \left(\eta R + 1-\eta \right) 
    \end{equation}
where $P_\text{in}$ is the power landing on the input mirror and $\eta$ denotes the fraction of the input beam that is mode-matched to the cavity resonance (the other fraction $1-\eta$ just reflects from the cavity, adding a constant offset to the collected power). Figure~\ref{fig:fig1}D shows the fractional collected power $P_\text{out}/P_\text{in}$ as a function of the total internal loss ${\mathcal{L}=\mathcal{L}_0+\mathcal{L}_\text{rad}}$ normalized by the input mirror transmission $T_1$ (teal curve). Transient increases in $\mathcal{L}_\text{rad}(t)$ (exaggerated black curves) modulate the reflected power (orange curves) in proportion to the slope $dP_\text{out}/d\mathcal{L}$ (of the red segments), producing a reflected power response that depends on the parameter regime of the cavity. Notably, for an ``overcoupled" cavity ($T_1 > \mathcal{L}_0$), hydrated electrons \textit{reduce} the reflected power, while for an ``undercoupled'' cavity ($T_1 < \mathcal{L}_0$), they \textit{increase} the reflected power. 

\begin{figure}
	\centering
	\includegraphics[width=.7\textwidth]{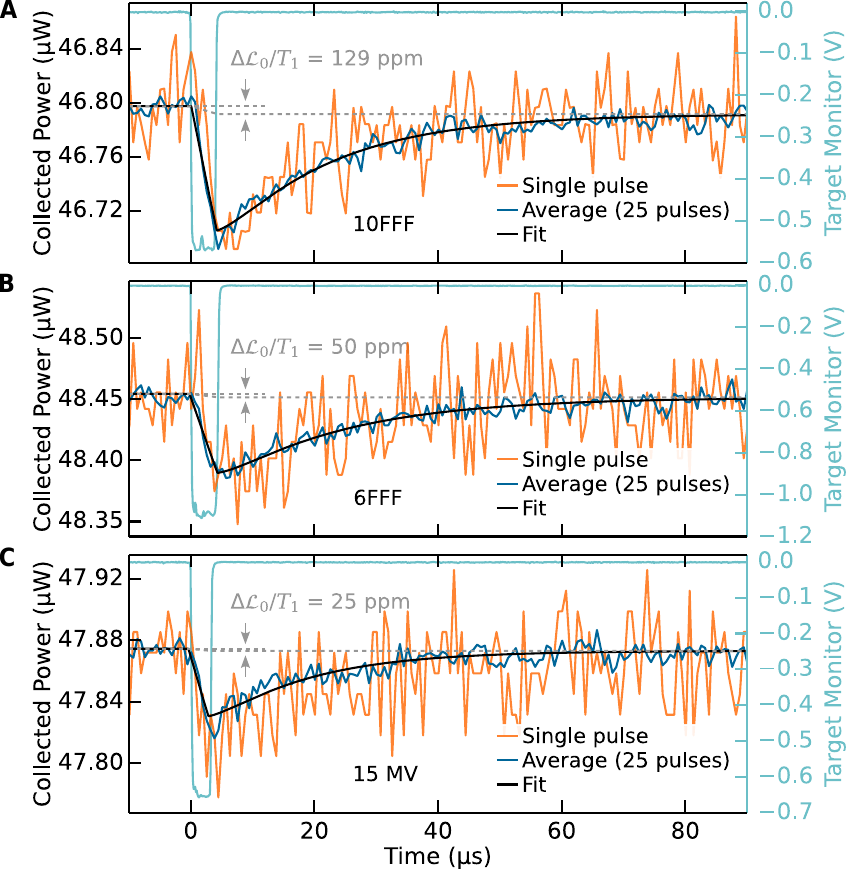} 
	\caption{\small
        \textbf{Single-shot readout of individual clinical radiation pulses.}
         Orange curves show the observed single-shot transients in the collected power (for ${P_\text{in}\approx0.8}$~mW sent to the cavity) for individual photon pulses with the LINAC set to (A) 10~MV, flattening filter free (FFF), nominally delivering 1.59~mGy/pulse to water, (B) 6~MV (FFF) delivering 0.93~mGy/pulse, and (C) 15~MV (flattened) delivering 0.71~mGy/pulse, all at 75~cm~SSD and $2.25\times2.25$~cm$^2$ field size centered on the basin's top window. As a proxy for delivered dose rate, the light teal curves show the LINAC target current monitor. The blue curve shows an average of 25 pulses, emphasizing the characteristic pulse shape (build-up during exposure followed by $\eaq$ decay) and repeatability.  
         Black curves show fits of averaged data to a model (main text) with fit parameters (A)~${D=1.57 \pm 0.03}$~mGy/pulse, ${\tau_\text{e}=17.7\pm 0.5\ \upmu}$s, (B)~${D=0.86 \pm 0.02}$~mGy/pulse, ${\tau_\text{e}=20.4\pm 0.9\ \upmu}$s, and (C)~${D=0.65 \pm 0.03}$~mGy/pulse, ${\tau_\text{e}=16\pm 1\ \upmu}$s. Gray dashed lines and arrows show the longer-time-scale increase $\Delta\mathcal{L}_0/T_1$ in background loss, attributed primarily to optical absorption by ozonide radicals.
        }
	\label{fig:fig2}
\end{figure}

Figure~\ref{fig:fig2} shows the observed transient changes in collected light associated with clinical doses of photon radiation for (A) 10~MV~flattening-filter-free (FFF) and 1.59~mGy/pulse, (B) 6~MV~FFF and 0.93~mGy/pulse, (C) 15~MV and 0.71~mGy/pulse. All three have a 75~cm source-to-surface distance (SSD), and a ${2.25\times2.25}$~cm$^2$ field size centered on the basin's top window. In all cases, we observe absorption transients from individual pulses (orange) exhibiting a characteristic build-up during exposure (i.e., within the LINAC's target monitor pulse (teal)) and subsequent decay expected from the radiation-induced $\eaq$ population \cite{keene1962,megroureche2023}. The blue trace shows an average of 25 such pulses acquired during exposure, emphasizing the reproducible shape of the transient. This demonstration of real-time, cavity-enhanced optical readout of individual, clinically relevant mGy pulses in a tissue-equivalent medium is the central result of our work.

\section*{Agreement with Coupled Rate Equations}

To quantitatively understand these signals, we model the time-dependent concentration $n_\text{e}$ ($n_\text{H}$)~[mol/L] of radiation-induced $\eaq$ (H) particles with the coupled rate equations
\begin{align}
\label{eq:H_rate}
    \partial_t n_\text{H}(t) &= \rho G_\text{H} d(t) - \frac{1}{\tau_\text{H}} n_\text{H}(t)\\
\label{eq:eaq_rate}
    \partial_t n_\text{e}(t) &= \rho G_\text{e} d(t) + \frac{1}{\tau_\text{H}} n_\text{H}(t) - \frac{1}{\tau_\text{e}} n_\text{e}(t),
\end{align}
where $\rho=0.997$~[kg/L] is the solution mass density at room temperature, $G_\text{H}=0.062$~$\upmu$mol/J ($G_\text{e}=0.28$~$\upmu$mol/J) is the radiation chemical yield for H ($\eaq$) particles~\cite{spinks1990}, $d(t)$~[Gy/s] is the dose rate from the LINAC, and $\tau_\text{H}$ ($\tau_\text{e}$)~[s] is the lifetime for H ($\eaq$) particles. Equation~\ref{eq:H_rate} includes the dominant decay of H particles via combination with \ce{OH^-} ions to produce $\eaq$ (and \ce{H_2O}), with lifetime ${\tau_\text{H}=3.4\ \upmu}$s for our pH~12.12 solution \cite{spinks1990}; consequently Eq.~\ref{eq:eaq_rate} includes the opposite as a secondary source term. Given the short, nearly rectangular pulses from the LINAC (teal curves in Fig.~\ref{fig:fig2}), we approximate $d(t)\approx D/\Delta t$ during exposure, where $D$~[Gy] is the dose per pulse and $\Delta t$~[s] is the pulse duration. Solving with initial condition $n_\text{e}(0) = n_\text{H}(0) = 0$ then yields a time-dependent $\eaq$ concentration
\begin{equation}\label{eq:ne_total}
n_\text{e}(t)=\begin{cases}
\left(\frac{\rho G_\text{e}D}{\Delta t}+\frac{\rho G_\text{H}D}{\Delta t}\right)\tau_\text{e}\left(1-e^{-t/\tau_\text{e}}\right)+\frac{\rho G_\text{H}D}{\Delta t}\frac{\tau_\text{e}\tau_\text{H}}{\tau_\text{e}-\tau_\text{H}}\left(e^{-t/\tau_\text{H}}-e^{-t/\tau_\text{e}}\right) & 0\le t\le\Delta t\\
\left[n_\text{e}(\Delta t)+\frac{\tau_\text{e}}{\tau_\text{e}-\tau_\text{H}}n_\text{H}(\Delta t)\right]e^{-\left(t-\Delta t\right)/\tau_\text{e}}-\frac{\tau_\text{e}}{\tau_\text{e}-\tau_\text{H}}n_\text{H}(\Delta t)e^{-\left(t-\Delta t\right)/\tau_\text{H}} & t>\Delta t
\end{cases}
\end{equation}
that is converted to a measurable signal via the corresponding optical absorption coefficient 
    \begin{align}       \label{eq:alpha_eaq}
    \alpha_\text{e}(t) &= \ln(10) \mathcal{E} n_\text{e}(t),
    \end{align}
(part of $\mathcal{L}_{\text{rad}}(t)$ in Eq.~\ref{eq:R}) where $\mathcal{E} ={1.76 \times 10^6}$~L mol$^{-1}$ m$^{-1}$ is the molar absorption coefficient for our wavelength 652~nm~\cite{torche2016}. 

We also consistently resolve a small, long-lived increase in background absorption $\Delta\mathcal{L}_0$ after each pulse that recovers on the millisecond time scale (see gray dashed lines). 
This step cannot be explained by heating, since the total $\sim$mGy dose delivered by each irradiation pulse can at most heat the solution by 0.2~$\upmu^\circ$C, which would change the background absorption by ${<1\times 10^{-12}}$~\cite{langford2001}; this is 7 orders of magnitude smaller than what is observed. Instead, we expect $\Delta\mathcal{L}_0$ arises from the ozonide radical \ce{O3^{.-}}, which is known to have a lifetime $\tau_{\text{oz}}$ of several~milliseconds~\cite{czapski1967}, and whose optical absorption tail likely extends to 652~nm~\cite{hug1981}. The large separation of time scales ($\tau_\text{e}\ll\tau_\text{oz}$) makes it easy to isolate the $\eaq$ contribution, as we can just include a background that increases $\mathcal{L}_0$ by $\Delta\mathcal{L}_0$ linearly during the pulse (since $\Delta t\ll \tau_{\text{oz}}$) and remains approximately constant over our measurement time after the pulse. 

As per Eq.~\ref{eq:Pout_and_eta}, the input power $P_\text{in}$ plays a crucial role in relating the observed output power $P_\text{out}$ to $\mathcal{L}_\text{rad}$ and dose $D$. As discussed in Methods, we perform many measurements like those of Fig.~\ref{fig:fig2} over 30 minutes, and can fit the data using the known dose $D$ to provide a reliable calibration value of $P_\text{in}$ (drifting) for each individual exposure. If we then assume the initial value $P_\text{in} = 0.8$ mW remains roughly constant over the first 5 minutes, we can instead fit for dose $D$ to show consistency in the relative values expected from our available beam parameters. These fits are shown in Fig.~\ref{fig:fig2} (black curves). The ratio of measured per-pulse dose~$D$ to the nominal dose~$D_0$ from each fit in Fig.~\ref{fig:fig2} is (A)~$0.99\pm0.07$ (calibrated value), (B)~$0.93\pm0.07$, and (C)~$0.92\pm0.07$. 

\begin{figure} 
	\centering
	\includegraphics[width=1\textwidth]{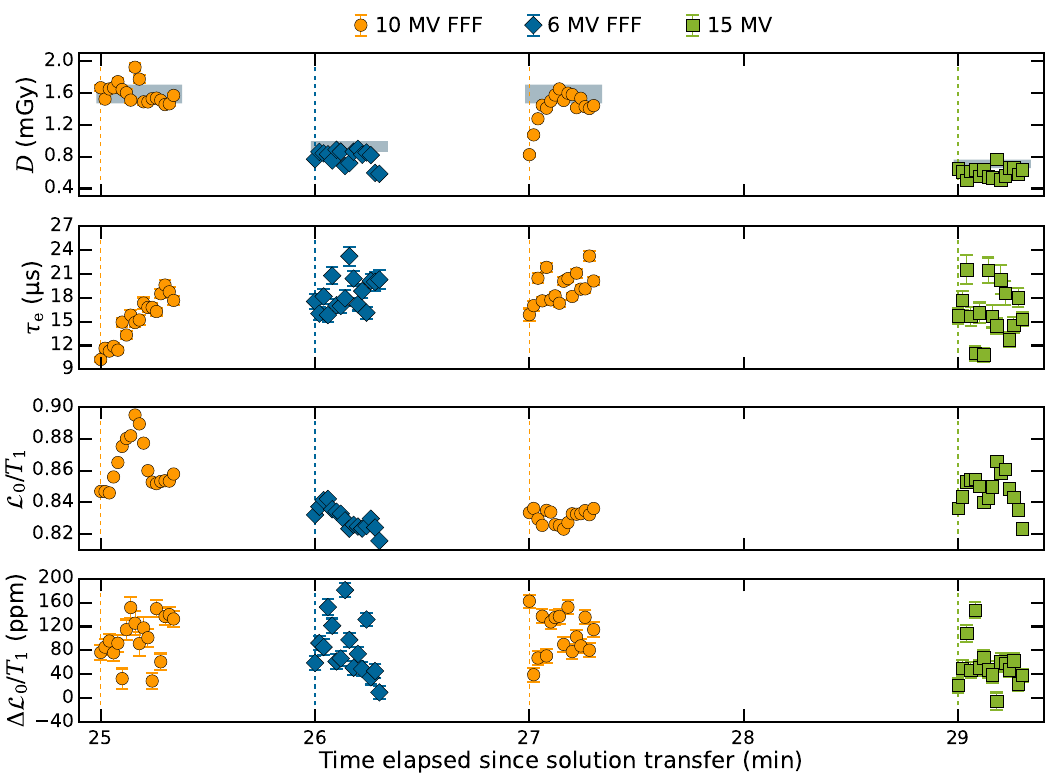} 
	\caption{
        \small
        \textbf{Time-evolution of fit parameters.}
		Inferred dose $D$ per pulse, $\eaq$ lifetime $\tau_\text{e}$, cavity loss ratio ${\mathcal{L}_0/T_1}$, and pulse-induced excess background loss ${\Delta\mathcal{L}_0/T_1}$ for 10~MV~FFF~(orange circles), 6~MV~FFF~(blue diamonds) and 15~MV~(green squares) photon beams. Time is measured relative to the moment of solution transfer. Each data point corresponds to a fit (black curves in Fig.~\ref{fig:fig2}) to an average of 25 irradiation pulses.
        Gray bars indicate the $\pm7.4\%$ systematic error (dominated by 5\% uncertainty in literature values of $G_\text{e}$, 3.6\% uncertainty in our film measurements, and $\sim$3\% uncertainty in cavity length $L$; see Methods) added to the nominal dose per pulse for each beam energy (1.59~mGy for 10~MV~FFF, 0.93~mGy for 6~MV~FFF, and 0.71~mGy for 15~MV) at the chosen irradiation field size~(${2.25 \times 2.25}$~cm$^2$ at the basin's surface), source-to-surface distance~(75~cm) and penetration depth~(2.5~cm).
        }
	\label{fig:fig3}
\end{figure}

Figure~\ref{fig:fig3} shows the time-evolution of all such fits over a $\sim$5-minute time scale, acquired approximately 25 minutes following transfer of the argon-purged NaOH solution into the cavity. Also shown is the range expected for our system parameters (gray shaded regions). Note we consistently see evidence that the irradiation itself affects the population of $\eaq$ scavengers: $\tau_\text{e}$ generally increases during initial irradiations~\cite{fielden1967}, then gradually decreases over subsequent hours (likely due to oxygen contamination). Promisingly, the dose inferred with this method is robust against $\sim$10\% drifts in $\mathcal{L}_0$ and the large changes in $\tau_\text{e}$ (notably the factor of 2 change starting at minute 25). This series also gives a sense of the system's drift and transients (e.g., minute 27). As discussed below, we believe this issue can be wholly mitigated by switching to a transmission modality.

\section*{Single-Pulse Sensitivity}

We can estimate the dosimeter's sensitivity to individual pulses in the absence of drifts by fixing all of the (now known) system parameters except $D$, and fitting the single-pulse transients (Fig.~\ref{fig:fig2} orange curves). The fit uncertainty on $D$ then yields sensitivities (A)~90~$\upmu$Gy for 10 MV photons, (B)~70~$\upmu$Gy for 6 MV photons, and (C)~80~$\upmu$Gy for 15 MV photons, consistent with the uncertainties of ${\pm20-30\ \upmu}$Gy from the aforementioned (less constrained) fits, since averaging over 25 traces should reduce the fluctuations and fit uncertainty by a factor of $\sim\sqrt{25}$. These sensitivity values essentially reflect the excess classical laser noise present in $P_\text{out}$ (which can be improved as discussed below) but are already significantly lower than the $\sim$mGy per-pulse doses delivered by the LINAC.

\section*{Mitigating Noise and Drift}

This work establishes a new paradigm for radiation dosimetry in which absorbed dose to water is measured directly in real time using an all‑optical, cavity‑enhanced readout of the induced hydrated electron population. By folding meter‑scale optical path lengths into a compact, water‑filled Fabry--Pérot cavity, we demonstrate single‑pulse sensitivity to clinically relevant mGy doses, notably resolving individual pulses from a medical linear accelerator with a tissue-equivalent sensing medium. These results highlight the potential of cavity‑enhanced radiolysis absorption spectroscopy as a universal dosimetry platform.

This prototype reveals several reasonable pathways to greatly improved performance and functionality. 
Presently, its sensitivity is limited by classical laser noise, which is not fundamental. Simple feedback techniques can eliminate classical noise in the relevant frequency range for $\sim$mW-scale input power~\cite{dumont2023}, in principle reducing the minimum detectable dose by at least an order of magnitude. 
We also find the inferred dose is quite sensitive to drifts in the input laser polarization (drifts in $P_\text{in}$) when operating in reflection mode. One could in principle reduce this issue with a combination of polarization-maintaining fibers and monitoring the input power with a pick-off before the cavity. Within the solution, we observe obvious drifts in the internal losses (sometimes the amplitudes of the transients even flip sign), consistent with the observed formation of microbubbles after irradiation. These, combined with thermal gradients may even change the cavity mode \textit{shape} and hence $\eta$, which may explain the remaining transients and drifts in the fit values of dose -- a subject of future study. However, these issues can in principle be fully mitigated by measuring cavity transmission instead of reflection (this requires a different set of cavity mirrors), wherein the detected power leaving the \textit{back} mirror provides a faithful record of the intracavity power (including drifts) without the added complication of interference between light leaving the cavity and the promptly reflected input beam (i.e., $\eta$). Transmission operation would therefore provide a self-calibrating fractional absorption measurement that is insensitive to drifts in power, cavity loss, and cavity mode shape, enabling stable long-term operation. 

Beyond improving readout robustness, future work will further establish the linearity and dynamic range of this detection scheme across modern and next-generation FLASH dose rates. With modest reductions in technical laser noise, it should be possible to systematically vary the delivered dose and directly compare the measured response with theoretical expectations based on radiolysis kinetics, providing a critical validation step.

The use of water as the sensing medium provides a unique opportunity to realize a dosimeter that operates consistently across all radiation types, energies, and dose rates. By directly measuring absorbed dose to water without needing material-dependent corrections, this approach could dramatically simplify quality assurance procedures -- notably redistributing absolute dosimetry from specialized labs to individual clinics -- and enable real-time verification in emerging modalities such as FLASH and MR-guided radiotherapy. Taken together, these advances point toward a new class of compact, tissue‑equivalent dosimeters capable of operating directly at the point of interest, with broad implications for clinical practice and radiation science.

%%%%%%%%%%%%%%%% MAIN TEXT FIGURES %%%%%%%%%%%%%%%

%%%%%%%%%%%%%%%% MAIN TEXT TABLES %%%%%%%%%%%%%%%

% \begin{table} % Do NOT use \begin{table*}
% 	\centering
% 	% Captions go above tables
% 	\caption{\textbf{All captions must start with a short bold sentence, acting as a title.}
% 		Then explain what is being listed in the table, the meaning of each column etc.
% 		Captions are placed above tables.}
% 	\label{tab:example} % give each table a logical label name
	
% 	\begin{tabular}{lccc} % four columns, alignment for each
% 		\\
% 		\hline
% 		Sample & $A$ & $B$ & $C$\\
% 		 & (unit) & (unit) & (unit)\\
% 		\hline
% 		First & 1 & 2 & 3\\
% 		Second & 4 & 6 & 8\\
% 		Third & 5 & 7 & 9\\
% 		\hline
% 	\end{tabular}
% \end{table}

%%%%%%%%%%%%%%%% REFERENCES %%%%%%%%%%%%%%%

\clearpage % Clear all remaining figures and tables then start a new page

%\bibliography{dosimeter}

\bibliographystyle{bibstyle}

% After the paper has completed peer review and been revised ready for acceptance,
% you should comment out the lines above and copy-paste the contents of your .bbl
% file here instead. This will help ensure that our conversion software works correctly.
% Remember to re-run BibTeX first - check the timestamp!
%
% Example of the first three entries copy-pasted from science_template.bbl:
%
%\begin{thebibliography}{1}
%
%\bibitem{example}
%A.~N. {Author}, An example reference. \emph{Journal of Improbable Research}
%  \textbf{1}, 67 (2020).
%
%\bibitem{example2}
%F.~M. {Surname}, S.~{Author}, A second example. \emph{Interesting Research
%  Letters} \textbf{32}, 897 (2019).
%
%\bibitem{example_preprint}
%P.~{One}, P.~{Two}, P.~{Three}, {An unpublished preprint}. \emph{preprint}
%  (2021), arXiv:2101.12345.
%
%\end{thebibliography}

%%%%%%%%%%%%%%%% ACKNOWLEDGEMENTS %%%%%%%%%%%%%%%

\section*{Acknowledgments}
We thank Hamed Bekerat and Robert Hopewell for helpful discussions and support during experiments.
% Here you can thank helpful colleagues who did not meet the journal's authorship criteria, or
% provide other acknowledgements that don't fit the (compulsory) subheadings below.
% Formatting requirements for each of these sections differ between the \textit{Science}-family
% journals; consult the instructions to authors on the journal website for full details.
%
\paragraph*{Funding:}
FCRM acknowledges support from the Fonds de Recherche du Québec -- Nature et Technologies (FRQNT) B2 scholarship (dossier \#2022-2023 - B2X - 319595). 
JCS acknowledges financial support from the Natural Sciences and Engineering Research Council of Canada (NSERC RGPIN 2018-05635 \& 2024-04620 \& 2023/05312, ALLRP 578464-22), the Canada Research Chairs (CRC 235060 \& 252135), the Canada Foundation for Innovation (CFI 228130, 36423), Institut Transdisciplinaire d'Information Quantique (INTRIQ), and the Centre for the Physics of Materials (CPM) at McGill.
LC acknowledges support from the Natural Science and Engineering Research Council of Canada (NSERC Discovery grant RGPIN 2020-04095), and the Canada Foundation for Innovation (Innovation Fund 2015 Project \#33488 and LOF/CRC 229003).
SAE acknowledges support from the Natural Sciences and Engineering Research Council of Canada (NSERC RGPIN 2023-05312), and the Canada Research Chairs (CRC 252135).
%
%List the grants, fellowships etc. that funded the research; use initials to specify which author(s) were supported by each source. Include grant numbers when appropriate or required by the funding agency.
% For example: F.~A. was funded by the Generous Science Agency grant~2372.
%
\paragraph*{Author contributions:}
JCS, LC and SAE conceived the project. 
FCRM led the research and coordinated the project under the supervision of JCS and LC. 
FCRM and SB designed and assembled the optical cavity apparatus. 
FCRM and KS performed cavity alignment and noise optimization. 
FCRM, JCS and KS performed the irradiation measurements.
JB prepared the aqueous solutions and helped prepare measurements.
JB and TC operated the LINAC. 
TC and KS performed the radiochromic film measurements for dose calibration.
FCRM developed the theoretical model, performed the calculations and data analysis, conducted the literature review, and prepared the figures.
FCRM, JCS and LC wrote the manuscript. 
All coauthors commented on the manuscript.

\paragraph*{Competing interests:}
There are no competing interests to declare.
\paragraph*{Data and materials availability:}
Data presented in this paper are open access and will be freely available at McGill's Dataverse \cite{2026quantum}.

\subsection*{Supplementary materials}
Materials and Methods\\
% Supplementary Text\\
Fig. S1
% References \textit{(7-\arabic{enumiv})}\\ % automatically fills out the last reference number
% (filling out the other numbers automatically is possible but fiddly and liable to break)

%%%%%%%%%%%%%%%% END OF MAIN TEXT %%%%%%%%%%%%%%%

\newpage

%%%%%%%%%%%%%%%% START OF SUPPLEMENT %%%%%%%%%%%%%%%

% Figures, tables, equations and pages in the supplement are numbered S1, S2 etc.
\renewcommand{\thefigure}{S\arabic{figure}}
\renewcommand{\thetable}{S\arabic{table}}
\renewcommand{\theequation}{S\arabic{equation}}
\renewcommand{\thepage}{S\arabic{page}}
\setcounter{figure}{0}
\setcounter{table}{0}
\setcounter{equation}{0}
\setcounter{page}{1} % not 0 as \newpage already started a supplementary page
% References continue the numbering from the main text.

%\input{methods.tex}
\section*{Materials and Methods}

The hydrated electron dosimeter is based on a free-space optical cavity configured for reflection-mode measurements. The cavity employs two 1''~diameter dielectric mirrors (Layertec 116284 and 140965) with nominal power reflectivity (96\% input, 99.998\% back) selected to maximize the outcoupled power from the input port (reflection mode) and to enhance sensitivity to small absorption changes in the aqueous sample. 
The mirrors are mechanically clamped to a polytetrafluoroethylene (PTFE) basin using a custom stainless-steel mount secured with stainless-steel screws to ensure mechanical stability. Although PTFE was used in this study, polyether ether ketone (PEEK) is recommended for future implementations because of its lower porosity and reduced risk of fluorine-containing gas production under ionizing radiation. 
The 3-cm-long custom-made basin has a 20~mm~diameter horizontal cylindrical cut sealed by the mirrors on each side, and an 18~mm~$\times$~20~mm top window for incoming radiation, for a total volume of $\sim15$~mL. The center of the mirrors (and cavity mode axis) lie $\sim$~25~mm below the basin top surface.

The cavity is interrogated using a diode laser (Toptica DL Pro) operated at a center wavelength of 652~nm. Shorter wavelengths would nominally increase the ratio of hydrated electron absorption to background absorption, but longer effective optical paths would then be needed to maintain sensitivity, requiring higher-finesse mirrors and complicating cavity locking. At the same time, mirror coating absorption and scattering increase at shorter wavelengths, increasing the intrinsic cavity loss~$\mathcal L_0$. The determination of an optimal probe wavelength therefore represents an open subject for future study.

The laser current is modulated at 22.57~MHz and delivered to the cavity via a 3.5~$\upmu$m-diameter-core single-mode optical fiber. The modulation primarily affects the laser frequency (rather than amplitude), producing sidebands for generation of the Pound-Drever-Hall error signal used to lock the cavity and laser on resonance~\cite{black2001}. The reflected cavity signal is picked up with a rotating wave plate and polarizing beam splitter, then coupled into a 400~$\upmu$m-diameter-core multimode fiber for delivery to a silicon photodiode (ThorLabs PDA8A) in the control room (away from radiation) and connected to a 20~MHz bandwidth oscilloscope (PicoScope 4824).

Deionized water (resistivity 18.2~M$\Omega\cdot$cm, Milli‑Q system, Millipore) adjusted to pH~12.12 by addition of 0.01~M~NaOH is used throughout. Raising the pH depletes the fast $\eaq$ scavenger H$^+$ and opens a feed-in pathway (\ce{H + OH^- \rightarrow e^-_{aq} + H_2O}) that partially converts the hydrogen atoms produced during radiolysis into hydrated electrons, further increasing the net observable $\left[\eaq\right]$~\cite{buxton1988,spinks1990} while not affecting the absorbed dose to water -- an energy‑per‑mass quantity (the corresponding NaOH mass fraction 0.04\% is negligible for dosimetric considerations). 
Prior to irradiation, 100~mL of the solution is bubbled with high-purity argon for 1–2~h in a sealed vial to remove dissolved oxygen. A 15~mL sample is subsequently transferred into the basin, which had been pre-purged with argon, and the assembly is sealed with Kapton tape to minimize oxygen influx during measurements.

Irradiations are performed using a clinical linear accelerator (Varian TrueBeam LINAC) with the top surface of the PTFE basin aligned 75~cm below the source. The collimator is set to $3\times3$~cm$^2$ at the LINAC isocenter (SSD~$=100$~cm), resulting in a $2.25\times2.25$~cm$^2$ field at the basin/Kapton surface to ensure full coverage of the top opening. Measurements were conducted at multiple beam energies over $\sim1$~h following transfer of the aqueous solution into the optical cavity. The LINAC was operated at the maximum available dose rate for each energy.
The nominal dose per pulse is calibrated with EBT3 radiochromic film (Ashland ISP Advanced Materials) dosimetry under the same collimator setting and SSD, with a 2.5~cm water-equivalent bolus serving as the build-up medium to simulate the optical cavity basin. The films are irradiated to a total of 126~MU at each beam energy and dose rate presented here, from which we extract the path-averaged output dose (Gy/MU) over the total optical cavity length $L=3$~cm.
Along the cavity length, the geometry consists of an 18~mm central opening -- with a 2.5~cm water column above the probing laser axis -- flanked by two 6~mm end segments, each comprising 1.5~cm of PTFE and 1~cm of water above the beam (Fig.~\ref{fig:fig1}). 
The deep-end segments (PTFE + water) sit at $\sim 3.8$~cm water-equivalent depth~\cite{NISTWebBook,ICRU44} and fall in the low-dose penumbra, so that the depth correction changes the line integral by $<1\%$.
We then find the single-pulse dose (Gy) by integrating the total area of the LINAC target monitor (V$\cdot$s; Fig.~\ref{fig:fig2}, teal) and comparing to a calibration measurement (MU/V$\cdot$s) with known total dose. 
The film calibration results in a 3.6\% systematic uncertainty in our estimated nominal dose ($\sim3\%$ reported film readout uncertainties in this dose range~\cite{wen2016,marroquin2016} and 2\% estimated from $\pm3$~mm central axis misalignment); the conversion from target monitor area to MU translates into a 1.1\% systematic error, coming from repeated measurements and an upper bound 1\% error in the LINAC MU counting.
Other cavity parameters are extracted by sweeping the cavity length through multiple longitudinal resonances to estimate cavity finesse $\mathcal{F}\approx69$, and the resonance peak depths then provide an estimate of the input mirror transmissivity $T_1=0.0546\pm0.0002$~\cite{rodrigues2022}.

The measured transients in reflected power are fit to Eq.~\ref{eq:Pout_and_eta} (using Eqs.~\ref{eq:R}, \ref{eq:ne_total}, and \ref{eq:alpha_eaq}) as follows.
\textit{Fixed parameters:} the solution density $\rho=0.997$~kg/L (water at room temperature, since the NaOH concentration is only $\sim0.04\%$); the radiation chemical yields ${G_\text{e}=0.28}$~$\upmu$mol/J and ${G_\text{H}=0.062}$~$\upmu$mol/J for the primary $\eaq$ and hydrogen-atom yields (each with a typical literature uncertainty of $\pm 5\%$~\cite{buxton1988}), the latter converting to $\eaq$ with time constant ${\tau_\text{H}=3.4\pm1.0\ \upmu}$s at our operating pH \cite{spinks1990}; the $\eaq$ molar absorption coefficient ${\mathcal{E}_{652\text{nm}}=(1.76\pm0.04)  \times 10^6}$~L~mol$^{-1}$~m$^{-1}$~\cite{torche2016}; the pulse duration ${\Delta t=(4.125, 4.5, 3.25, 3.75)\ \upmu}$s -- estimated from the LINAC target current monitor for (10~MV~FFF, 6~MV~FFF, 15~MV, 10~MV) respectively; and the cavity length $L=30\pm1$~mm.
\textit{Cavity coupling:} We determine the cavity coupling $\eta$ by fitting the last 10~MV~FFF dataset (Fig.~\ref{fig:fig3}) with the calibrated dose $D$~[Gy].
We measure the off-resonance reflected power $P_\text{in}'$ as a proxy for $P_\text{in}$ immediately after the last irradiation in the series reported below, and assume $P_\text{in}$ did not have time to drift significantly. 
This fit gives ${\eta=0.9213 \pm 0.0001}$, $\eaq$ lifetime ${\tau_\text{e}=27.6\pm0.8\ \upmu}$s, cavity loss ratio ${\mathcal{L}_0/T_1=0.867 \pm 0.001}$, and transient excess background loss ${\Delta\mathcal{L}_0/T_1=220\pm20}$~ppm. 
This value of $\eta$ is consistent with an upper bound $\eta< 0.9490 \pm 0.0005$, obtained by sweeping the cavity length and fitting the amplitudes of all the visible higher-order transverse modes. It also agrees to within 0.1\% with $\eta=0.9224$ estimated 3~minutes before the first reported exposure -- a moment when $\mathcal{L}_0$ drifted through critical coupling ($\mathcal{L}=T_1$), evidenced by an inversion of the transient signal (as per the teal curve in Fig.~\ref{fig:fig1}D). This crossing allows a direct measurement of the depth $1-\eta$ of the resonance dip (teal curve minimum; accounting for the drifted value of $P_\text{in}$ discussed below). We therefore adopt $\eta=0.9213$ for all subsequent fits.

\begin{figure} 
	\centering
	\includegraphics[width=1\textwidth]{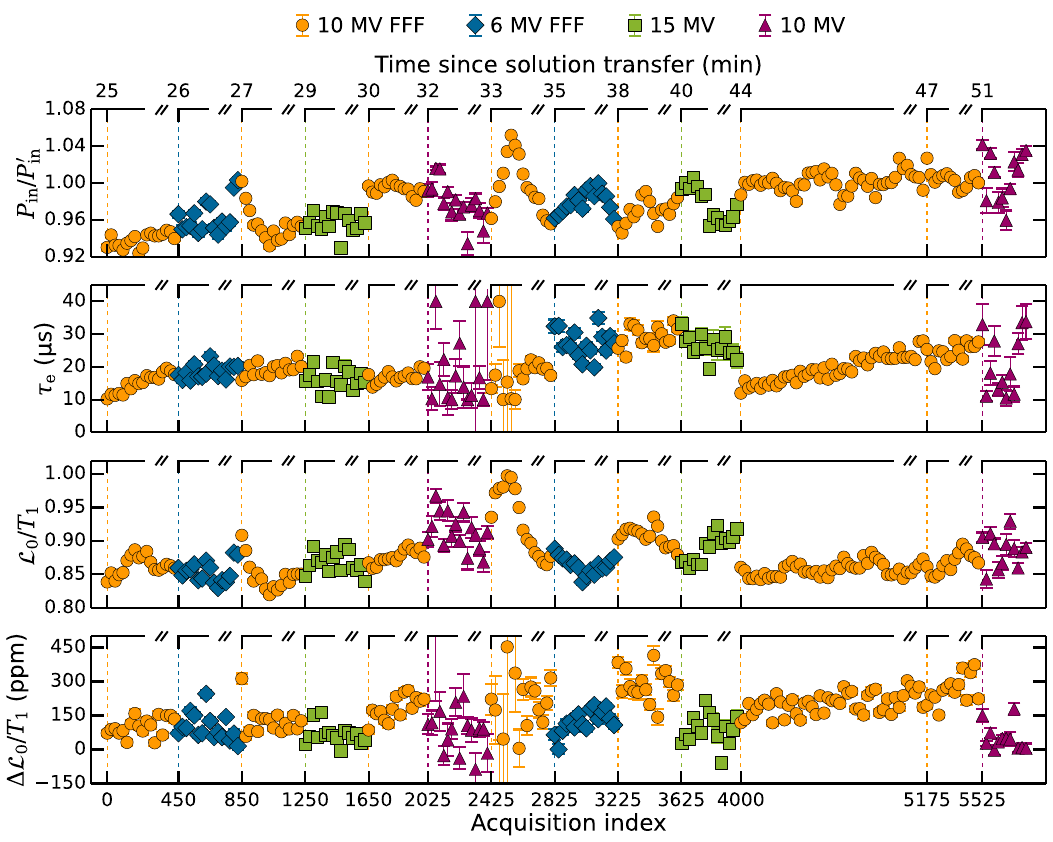} 
	\caption{
        \small
        \textbf{Time-evolution of cavity parameters.}
        Cavity input power $P_\text{in}$ (normalized to post-irradiation off-resonance power $P_\text{in}'$ measured at minute 54), $\eaq$ lifetime $\tau_\text{e}$, cavity loss ratio $\mathcal{L}_0/T_1$, and pulse-induced excess background loss $\Delta{\mathcal{L}_0/T_1}$, during irradiation runs over $\sim$~1~h. The dosimeter was exposed to 10~MV~FFF~(orange circles, 1.59~mGy/pulse), 6~MV~FFF~(blue diamonds, 0.93~mGy/pulse), 15~MV~(green squares, 0.71~mGy/pulse) and 10~MV~(purple triangles, 0.35~mGy/pulse) photon beams. Time is measured relative to the moment of solution transfer. Each data point corresponds to a fit (black curves in Fig.~\ref{fig:fig2}) to an average of 25 irradiation pulses. Field settings: ${2.25 \times 2.25}$~cm$^2$ irradiation field size (at basin's top surface), 75~cm source-to-surface distance, and 2.5~cm penetration depth.
        }
	\label{fig:sup1}
\end{figure}

The extracted dose therefore depends on the set of optical, geometric, and physicochemical parameters described above. When $P_\text{in}$ is well calibrated, the statistical uncertainty from fitting the transient response is on the order of ${2-3\%}$ (Fig.~\ref{fig:fig2}). The additional systematic contributions from the multiple parameters and dose calibration result in a $\pm7.4\%$ uncertainty in $D$, which is represented by the gray bars in Fig.~\ref{fig:fig3}.
% INPUT POWER DRIFT
Polarization (especially in long fibers) combined with laser drifts is known to produce $P_\text{in}$ fluctuations of $\sim10\%$ over the scale of tens of minutes in our system. Since the laser is locked to resonance during exposures, we leave $P_\text{in}$ as a fit parameter, observing the expected $\sim 10\%$ drifts shown in Fig.~\ref{fig:sup1}.

As mentioned in the main text, we observe drifts in the average reflected power between the irradiation runs from laser and polarization drifts (in our long fiber), and from changes in $\mathcal{L}_0$. The latter are consistent with microbubble formation within the solution, which can be observed at the internal surfaces of solution vials and cavity mirrors after irradiation. We expect they introduce additional scattering and refraction losses along the optical path, effectively increasing $\mathcal{L}_0$ in a manner that can vary on the few-seconds timescale. Because $\mathcal{L}_0/T_1$ is a direct fit parameter, however, such loss fluctuations can in principle be quantitatively tracked and corrected for in the analysis.
Finally, the observed absorption tail attributed to the ozonide radical is shown in Fig.~\ref{fig:sup1} as the added background loss $\Delta\mathcal{L}_0/T_1$.
There still exist transient events, however, that could be due to secondary effects (e.g., these solution dynamics somehow change the cavity mode shape and $\eta$) that warrant further investigation. We also note that in a cavity system optimized to collect light from the back mirror (transmission mode), the signal would have no dependence on $\eta$, greatly simplifying the interpretation in the presence of these drifts.

\end{document}